\documentclass[12pt]{article}
\usepackage{graphicx}
\usepackage{amsmath,latexsym,epsfig,epsf,rotate,amssymb}
\newcommand{\be}{\begin{equation}}
\newcommand{\ee}{\end{equation}}
\begin{document}

\title{Final state radiation and a possibility to
test a pion-photon interaction model near two-pion threshold. }

\author{G.~Pancheri$^{1)\footnote{e-mail: Giulia.Pancheri@lnf.infn.it}}$
, O.~Shekhovtsova$^{1),2)\footnote{e-mail:
shekhovtsova@kipt.kharkov.ua}}$,
 G.~Venanzoni$^{1)\footnote{e-mail: Graziano.Venanzoni@lnf.infn.it}}$
\\
\\
\emph{$^{1)}$INFN Laboratori Nazionale di Frascati, Frascati (RM) 00044, Italy} \\
\emph{$^{2)}$NSC ``Kharkov Institute for Physics and Technology'',} \\
\emph{Institute for Theoretical Physics, Kharkov 61108, Ukraine }
}
\date{}

\maketitle

\begin{abstract}
Final state radiation in the process $e^+e^-\to\pi^+\pi^-$
is considered for the cuts used in 
the analysis of KLOE data at large angles. By means of a Monte Carlo event generator FEVA, effects of  non-pointlike behaviour of pions are estimated in the
framework of Resonance Perturbation Theory. An additional
complication related with the $\phi$ meson intermediate state is
taken into account and the corresponding contributions (the direct
decay $\phi\to\pi^+\pi^-\gamma$ and the double resonance decay
$\phi\to\rho^\pm\pi^\mp\to\pi^+\pi^-\gamma$) are added to  FEVA.
A method to test effects of non-pointlike behaviour of pions
in a model-independent way is proposed.
\end{abstract}

\vspace{0.5cm} PACS: 13.25.Jx, 12.39.Fe, 13.40.Gp
\section{Introduction}
\label{intro}

The ongoing experiments on precise measurements of the cross
section of $e^+e^-$ annihilation into hadrons aim to 
a precision at the $0.5-1.0\%$ level \cite{cmd2,kloe,snd}.
Such an accuracy is crucial for various tests of the
Standard Model \cite{eid_jeg}, e.g., by confronting the experimentally measured
value of $a_\mu$ \cite{Upgm}, the muon anomalous magnetic moment,
 to the theoretical prediction. The accuracy of the
theoretical calculation of $a_\mu$ is currently limited by the
hadronic contribution, $a_\mu^{(had)}$. This contribution cannot
be reliably calculated in the framework of perturbative QCD,
because low-energy region dominates.  Fortunately, its leading order part, $a_\mu^{(had; LO)}$, can be
estimated from the dispersion relation using the experimental
cross sections of $e^+e^-$ annihilation as an input
\cite{eid_jeg}
\begin{equation}\label{dispers}
a_\mu^{(had; LO)}=\Bigl(\frac{\alpha m_\mu}{3\pi}\Bigr)^2\int_{4
m_\pi^2}^\infty\frac{R(q^2)K(q^2)}{q^4}dq^2 , \; \; \; \;
R(q^2)=\frac{\sigma_h(q^2)}{\sigma_\mu(q^2)} .
\end{equation}
where $\sigma_h(q^2)$ is the total hadronic cross section,
$\sigma_\mu(q^2)$ is the total cross section of the process
$e^+e^-\to\gamma\to\mu^+\mu^-$ and
the function $K(q^2)$ is a smooth function that increases from
$0.63$ at the threshold ($s=4m_\pi^2$) upto $1$ for
$q^2\to\infty$. The quantity $q^2$ is the total four momentum squared of the final hadrons. 

This behaviour of the integrand results in that the biggest
contribution (about $70\%$) to the leading order hadronic part of the muon anomalous
magnetic moment, $a_\mu^{(had; LO)}$, comes from the energy region
below or about $1$ GeV. Due to the presence of the $\rho$-meson
the main contribution is related with the $\pi^+\pi^-$ final
state.

Experimentally, the energy region from threshold to the collider
beam energy is explored at the $\Phi$-factory DA$\Phi$NE
(Frascati, $s=4E^2=m_\phi^2$) \cite{Cataldi_99,binner} and
B-factory, PEP-2 (SLAC, $s=m_{\Upsilon(4S)}^2$) and KEKB (KEK,
$s=m_{\Upsilon(4S)}^2$) \cite{Sol,Benayoun_99} using the method of
radiative return \cite
{Chen_75,Rr1,Rr2}. 
This method  relies on a factorization of the radiative cross
section  in the product of the hadronic cross section and a radiation
function $H(q^2,\theta_{max},\theta_{min})$ known from Quantum
Electrodynamics (QED) \cite{Rr2,Baier_65,Khoze_02}.
The large luminosity of the $\Phi$ and
$B$-factories allows to compensate
 the additional factor $\displaystyle\frac{\mathstrut \alpha}{2\pi}$ caused by the hard photon emission.

For a two-pion final state considered here  it means
that the radiative cross section $\sigma^{\pi\pi\gamma}$
corresponding to the process
\begin{equation}\label{process}
e^+(p_1)+e^-(p_2)\to\pi^+(p_+)+\pi^-(p_-)+\gamma(k) ,
\end{equation}
can be written as
\begin{equation}\label{factoriz}
q^2\frac{d\sigma^{\pi\pi\gamma}}{dq^2}=\sigma^{\pi\pi}(q^2)H(q^2,\theta_{max},\theta_{min}) , \; \; \; q=p_++p_- , 
\end{equation}
where the hadronic cross section $\sigma^{\pi\pi}$ is taken at a
reduced CM energy.
This factorization is valid only for photon radiation from the
initial leptons (initial state radiation (ISR)). This is not possible for final
state (FS) radiation (FSR) which is an irreducible background in
radiative return measurements of  the hadronic cross section
\cite{kloe,czyz_old}.
~\footnote{In
fact the process of FSR cannot be excluded from the analysis. It
can be suppressed choosing the small angle kinematics
($\theta_\gamma<\theta_{max}\ll 1$) but for the large angle
analysis this contribution becomes upto $40\%$ of ISR and should
be estimated very carefully. (For some advantages of large angle
analysis compared to small angle one see \cite{kloe_large}.)} Indeed,
the FSR cross section calculation has an additional complication
compared to ISR case. In principle, RC caused by initial state
radiation, i.e. the function $H(q^2,\theta_{max},\theta_{min})$,
can be calculated in QED, although the accuracy
 is technically limited. Instead, for the calculation of the FSR cross section
the situation is different since its evaluation relies
on models describing pion--photon interaction. Usually the
combined sQED$*$VMD model is assumed as a model to calculate FS
Bremsstrahlung process \cite{Rr2, phokhara}. In this case the
pions are treated as point-like particles (the sQED model) and the
total FSR amplitude is multiplied by the pion form factor, that is
estimated in the VMD model. Unfortunately, the sQED$*$VMD model is
an approximation that is valid for relatively soft photons and it
can fail for high energy photons,
 i.e near the $\pi^+\pi^-$ threshold.
In this energy region the contributions to  FSR, beyond the
sQED$*$VMD model, can become important. As it was shown in
Ref.~\cite{Ecker_89} the Resonance Perturbation Theory (RPT) is an
appropriate model to describe photon-meson interactions in the
energy region about $1$ GeV and we will apply this model to estimate the Bremmstrahlung FS contributions beyond of sQED.

At thet $\Phi$ factory DA$\Phi$NE there is an additional complication
related with the  possible intermediate $\phi$ meson state and the
corresponding contributions should be included in the Monte Carlo event
generator.

In this paper we present the results obtained by the Monte Carlo
event generator FEVA that simulates the process (\ref{process})
for the  DA$\Phi$NE accelerator setup. Our
computer code FEVA was inspired by MC EVA \cite{Rr2}. The
previous version of FEVA was described in~\cite{our_phys}, where
the Bremsstrahlung process (both in the framework of RPT and sQED)
and the $\phi$ direct decay (only $f_0$ parameterization) were
considered. The current version of FEVA includes in addition the
double resonance contribution and a more sophisticated
parameterization for the $\phi$ direct decay.

This paper is organized as follows. In Section~\ref{fsr_model} we
give a general description of FSR process and present the FSR
models that have been already included in our program FEVA.  In
Section~\ref{numeric} 
the numerical results for the KLOE large angle analysis are presented. Since
 most of the effects arising
in the FSR are model-dependent, we conclude this paper by
suggesting a way 
 to test possible effects
beyond sQED, in a model-independent way (see Section
\ref{mod_ind}). A conclusion is given in Section~\ref{conclusion}.

\section{Final state radiation models}\label{fsr_model}

The cross section of the FSR  process can be written as
\begin{equation}\label{sect_fsr}
d\sigma_{F}=\frac{1}{2s(2\pi)^5}\int
\delta^4(Q-p_+-p_--k)\frac{d^3p_+d^3p_-d^3k}{8E_+E_-\omega}|M^{(FSR)}|^2
,
\end{equation}
where $Q=p_1+p_2$, $s=Q^2$,
\begin{equation}
M^{(FSR)}=\frac{e}{s}M^{\mu\nu}\bar u(-p_1)\gamma_\mu
u(p_2)\epsilon^\star_{\nu} ,
\end{equation}
and where the FS tensor $M^{\mu\nu}$  describes the transition
$$\gamma^*(Q)\to\pi^+(p_+)\pi^-(p_-)\gamma(k) .$$  It is convenient to parametrize the FS tensor in terms
of three gauge invariant tensors (see \cite{our_2005} and Ref.
$[23,24]$ in it):
\begin{eqnarray}\label{eqn:fsr}
&&M^{\mu \nu }(Q,k,l)\equiv -ie^{2}M_{F}^{\mu \nu }(Q,k,l)
=-ie^{2}(\tau _{1}^{\mu \nu }f_{1}+\tau
_{2}^{\mu \nu }f_{2}+\tau _{3}^{\mu \nu }f_{3}) ,    \\
&&\tau _{1}^{\mu \nu }=k^{\mu }Q^{\nu }-g^{\mu \nu }k\cdot Q, \;  \; \; \; \; l=p_+-p_- , \nonumber \\
&&\tau _{2}^{\mu \nu }=k\cdot l(l^{\mu }Q^{\nu }-g^{\mu \nu
}k\cdot
l)+l^{\nu }(k^{\mu }k \cdot l-l^{\mu }k \cdot Q) , \;  \nonumber \\
&&\tau _{3}^{\mu \nu }=Q^{2}(g^{\mu \nu }k\cdot l-k^{\mu }l^{\nu
})+Q^{\mu }(l^{\nu }k\cdot Q-Q^{\nu }k\cdot l) . \nonumber
\end{eqnarray}
We would like to emphasize that this expansion is totally
model-independent. The model dependence is related only with an
explicit form of the scalar functions $f_i$ (we will call them
structure functions).

Here is the list of the  FSR processes included in FEVA MC:
\begin{eqnarray}\label{fsr_proc}
e^++e^-&\to&\pi^++\pi^-+\gamma
\text{\hspace{5.5cm}\textbf{Bremsstrahlung process}}
\\
e^++e^-&\to&\phi\to (f_0;f_0+\sigma)\gamma\to\pi^++\pi^-+\gamma
\label{phi_direct} \text{\hspace{3.8cm}\textbf{$\phi$ direct
decay}}
\\
e^++e^-&\to&\phi\to \rho^\pm\pi^\mp\to\pi^++\pi^-+\gamma
\label{phi_vmd} \text{\hspace{4.1cm}\textbf{VMD contribution}} \\
e^++e^-&\to&\rho^\pm\pi^\mp\to\pi^++\pi^-+\gamma  .
\end{eqnarray}
In the next sections we present the models describing these
processes. The presence of (\ref{phi_direct}) and (\ref{phi_vmd})
 is due to  the energy at which KLOE is running($s=m_\phi^2$). 

\subsection{Bremsstrahlung process}\label{Br_proc}

As it was mentioned in Introduction, the sQED$*$VMD model is an approximation to describe soft photon radiation by pions.
To estimate the
contributions beyond the sQED$*$VMD model we applied RPT. The
model is based on Chiral Perturbation Theory ($\chi$PT) with the
explicit inclusion of the vector and axial--vector mesons,
$\rho_0(770)$ and $a_1(1260)$. Whereas $\chi$PT gives correct
predictions on the pion form factor at very low energy, RPT is the
appropriate framework to describe the pion form factor at intermediate energies ($E
\sim m_\rho$)~\cite{Ecker_89}~\footnote{In that paper it was shown
that the coupling constants of the effective chiral lagrangian at
the order $p^4$ are essentially saturated by meson resonance
exchange.} and
satisfies the  QCD high energy behaviour.

Using the result of Ref.~\cite{our_2005} we write the contribution
to the functions $f_i$  (see Eq.(\ref{eqn:fsr})) caused by the
Bremsstrahlung FS process as
\begin{eqnarray}\label{f}
f_{i}&=&f_{i}^{sQED}+\Delta f^{RPT}_{i}, \\
f_1^{sQED}&=&\frac{2k\cdot Q F_\pi(Q^2)}{(k\cdot Q)^2-(k\cdot
l)^2}, \; \; \; f_2^{sQED}=\frac{-2 F_\pi(Q^2)}{(k\cdot
Q)^2-(k\cdot l)^2}, \;\; \\
f_3^{sQED}&=&0 ,
\end{eqnarray}
where
\begin{eqnarray}\label{d_f}
\Delta f^{RPT}_{1} &=&\frac{F_{V}^{2}-2F_{V}G_{V}}{f_{\pi }^{2}}\biggl(\frac{1}{%
m_{\rho }^{2}}+\frac{1}{m_{\rho }^{2}- s-\mathrm{i}m_\rho\Gamma_\rho(s)}\biggr)  \nonumber \\
&-&\frac{F_{A}^{2}}{f_{\pi }^{2}m_{a}^{2}}\biggl[ 2+\frac{(k\cdot l)^{2}}{%
D(l)D(-l)}+\frac{(s+k\cdot Q)[4m_{a}^{2}-(s+l^{2}+2k\cdot Q)] }{
8D(l)D(-l)}\biggr],  \label{eq:delta-f1} \\
\Delta f^{RPT}_{2} &=&-\frac{F_{A}^{2}}{f_{\pi }^{2}m_{a}^{2}}\frac{%
4m_{a}^{2}-(s+l^{2}+2k\cdot Q)}{8D(l)D(-l)} ,  \; \; \; l=p_+-p_- , \label{eq:delta-f2} \\
\Delta f^{RPT}_{3} &=&\frac{F_{A}^{2}}{f_{\pi }^{2}m_{a}^{2}}\frac{k\cdot l}{%
2D(l)D(-l)} , \; \; \; D(l)=m_a^2-(s+l^2+2kQ+4kl)/4 .
\end{eqnarray}

For notations and details of the calculation  we refer a reader
to~\cite{our_2005}. $F_V$, $G_V$ and $F_A$ are parameters of the
model.  According to the RPT model the pion form factor, that includes $\rho-\omega$ mixing, can be written  as:
\begin{equation}\label{formfact}
F_\pi(q^2)=1+\frac{F_V G_V}{f_\pi^2}B_\rho(q^2)
\Biggl(1-\frac{\Pi_{\rho\omega}}{3q^2}B_\omega(q^2)\Biggr) ,
\end{equation}
where
\begin{equation}\label{rho}
B_{r}(q^2)=\frac{q^2}{m_r^2-q^2-\mathrm i m_r
\Gamma_r(q^2)} ,
\end{equation}
$q^2$ is the virtuality of the photon, $f_\pi=92.4$ MeV  and the
parameter $\Pi_{\rho\omega}$ describes $\rho$-$\omega$ mixing. An
energy-dependent width  is considered for the  $\rho$ meson:
\begin{equation}\label{gamma}
\Gamma_\rho(q^2)=\Gamma_\rho\sqrt{\frac{m_\rho^2}{q^2}}
\Biggl(\frac{q^2-4m_\pi^2}{m_\rho^2-4m_\pi^2}\Biggr)^{3/2}\cdot \Theta(q^2-4m_\pi^2),
\end{equation} while for
the $\omega$--meson a constant width is used, $\Gamma_\omega=8.68$
MeV, and $m_\omega=782.7$ MeV. We assume that the parameter
$\Pi_{\rho\omega}$, that determines $\rho$-$\omega$ mixing, is a
constant and is related to the branching fraction
$Br(\omega\to\pi^+\pi^-)$:
\begin{equation}\label{br_om}
Br(\omega\to\pi^+\pi^-)=\displaystyle\frac{\mathstrut
|\Pi_{\rho\omega}|^2}{\Gamma_\rho \Gamma_\omega m_\rho^2} .
\end{equation}
The value of  $F_V$ and $G_V$ as well as the mass of the $\rho$ meson ($m_\rho$) and the parameter of the $\rho$-$\omega$ mixing $\Pi_{\rho\omega}$
were estimated by the fit of Novosibirsk CMD-2 data for the pion form factor \cite{cmd2}:
\begin{eqnarray*}\label{num_with}
&& m_\rho=774.97\pm 1.4 \text{ MeV} , \; \; \; \Pi_{\rho\omega}=-2774\pm 291.2 \text{ MeV}^2 ,
\nonumber \\&& \Gamma_\rho=145.21\pm2.6 \text{ MeV} , \; \; \; F_V=154.22\pm0.5 \text{ MeV} .
\end{eqnarray*}
Then  $G_V=64.6\pm0.3$ MeV and
$Br(\omega\to\pi^+\pi^-)=(0.96\pm0.19) \%$.

For the $a_1$ meson  we take $m_a=1.23$ GeV and $F_A=0.122$ GeV corresponding to the mean value of the experimental decay width $\Gamma(a_1\to\pi\gamma)=640\pm 246$ keV \cite{pdg}.

We would like to mention here that the contribution of any model
describing the Bremsstrahlung FS process can be conveniently rewritten as in Eq.(\ref{f}) and
in the soft photon limit the results should coincide with the sQED$*$VMD model
prediction.

\subsection{$\phi$ direct decay}

For DA$\Phi$NE energy ($s=m_\phi^2$) there are contributions
to the final state $\pi^+\pi^-\gamma$ related with the
intermediate $\phi$ meson state. In this section we consider the
direct rare decay $\phi\to \pi^+\pi^-\gamma$.

The $\phi$ direct decay is assumed to proceed through the 
intermediate scalar meson state (either $f_0$ or $f_0+\sigma$):
 $\phi\to (f_0;f_0+\sigma)\gamma\to\pi^+\pi^-\gamma$,  and its mechanism is
described by a single form factor $f_\phi(Q^2)$.
As it was shown
in~\cite{graz,gino}, this process affects the form factor $f_1$ of
Eq.({\ref{eqn:fsr}}):
\begin{equation}\label{phi_ampl}
f_1^{scal}=\frac{g_{\phi\gamma}f_\phi(Q^2)}{s-m_\phi^2+im_\phi\Gamma_\phi} .
\end{equation}

First, we consider the case of the $f_0$ intermediate state. To
estimate this contribution we apply the Achasov four-quark model
described in~\cite{ach_sol}: the $\phi \to f_0 \gamma$ decay
amplitude is generated dynamically through the loop of charged
kaons. The form factor $f_\phi$ reads:
\begin{equation}\label{fphi}
f_\phi^{K^+ K^-}(Q^2) = \frac{g_{\phi K^+ K^-}}{2\pi^2 m^2_{K}}\frac{g_{f_0 \pi^+
\pi^-} g_{f_0 K^+ K^-}e^{i\delta_B(Q^2)}}{(m_{f_0}^2 -Q^2
+Re\Pi_{f_0}(m_{f_0}^2) - \Pi_{f_0}(Q^2))} I\left ( \frac
{m_\phi^2}{m_K^2},\frac {Q^2}{m_K^2} \right )  ,
\end{equation}
where $I(.,.)$ is a function known in analytic
form~\cite{graz,close} and $\delta_B(Q^2) = b\sqrt{Q^2-4
m_{\pi}^2} $, $b=75^o/$GeV. The term
$Re\Pi_{f_0}(m_{f_0}^2) - \Pi_{f_0}(Q^2)$ takes into account the
finite width corrections to the $f_0$ propagator~\cite{ach_sol}. A
fit to the KLOE data
$\phi\to\pi^0\pi^0\gamma$~\footnote{$\Gamma(f_0\to\pi^+\pi^-)=
\frac{2}{3} \Gamma(f_0\to\pi\pi)$}  gives the following values of
the parameters~\cite{kloepi0_old}:
\begin{equation}
m_{f_0}= 0.962~{\rm GeV}, \,\, \frac{g^2_{f_0 K^+ K^-}}{4\pi} =
1.29~{\rm GeV}^2, \,\, \frac{g^2_{f_0 K^+ K^-}}{g^2_{f_0 \pi^+
\pi^-}} = 3.22.
\end{equation}

In a refined version of this model which includes the
$\sigma$ meson in the intermediate
state~\cite{achasov}, the form factor $f_\phi$ can be written as
\begin{equation*}
f_\phi^{K^+ K^-}(Q^2)=\frac{g_{\phi K^+ K^-}}{2\pi^2 m^2_{K}}e^{i(\delta_{\pi\pi}(Q^2)+\delta_{KK}(Q^2))}I\left ( \frac
{m_\phi^2}{m_K^2},\frac {Q^2}{m_K^2}\right ) \sum_{R,R'}g_{RK^+K^-}G^{-1}_{RR'}g_{R'\pi^+\pi^-} , 
\end{equation*}
where $G_{RR'}$ is the matrix of inverse propagators \cite{achasov}. 
Such an extension of the model improves the
description of the data at low $Q^2$ (see Fig. 1, {\it left}) and
gives the following value of the model parameters \cite{kloepi0}
\begin{eqnarray}
m_{f_0}&=& 0.977~{\rm GeV} , \,\, \frac{g^2_{f_0 K^+ K^-}}{4\pi} =
1.12~{\rm GeV}^2 , \,\, \frac{g^2_{f_0 K^+ K^-}}{g^2_{f_0 \pi^+
\pi^-}} =6.9  , \\
m_{\sigma}&=& 0.462~{\rm GeV} , \,\, \frac{g^2_{\sigma K^+
K^-}}{4\pi} = 0.024~{\rm GeV}^2 , \,\, \frac{g^2_{\sigma K^+
K^-}}{g^2_{\sigma \pi^+ \pi^-}} = 0.052 .
\end{eqnarray}

\subsection{VMD contribution}

Another contribution producing the intermediate $\phi$ meson state
is the double resonance contribution (\ref{phi_vmd}). In this case the off-shell $\phi$ meson
decays to $\rho^\pm$ and $\pi^\mp$ followed by $\rho\to\pi\gamma$.
The explicit value for the functions $f_i^{VMD}$ for this decay
can be found in Ref. \cite{gino}. To correspond to the KLOE
analysis \cite{kloepi0} we added also the additional phase between
VMD and $\phi$  direct contributions, the factor
$\Pi_\rho^{VMD}$~\footnote{Including $\Pi_\rho^{VMD}$ one rescales
the coupling constant. In our opinion it rescales the constant
$g^\phi_{\rho\pi}$ that cannot be directly determined from any
experimental decay width.} and the phase of the $\omega$-$\phi$
meson mixing $\beta_{\omega\phi}$:
\begin{eqnarray}
f_1^{VMD}&=& -\frac{1}{4\pi\alpha
s}\Bigl((-1+\frac{3}{2}x+\sigma)(g(x_1)+g(x_2)) \nonumber \\
&&+\frac{1}{4}(x_1-x_2)(g(x_1)-g(x_2))\Bigr)  , \\
f_2^{VMD}&=&-\frac{1}{4\pi\alpha s^2}(g(x_1)+g(x_2))  , \nonumber \\
f_3^{VMD}&=&-\frac{1}{8\pi\alpha s^2}(g(x_1)-g(x_2)) , \nonumber
\end{eqnarray}
where
\begin{equation}
g(x)=\frac{e
g^\phi_{\rho\pi}g^\rho_{\pi\gamma}}{4F_\phi}\frac{m_\phi^2 e^{\rm
i \beta_\rho}e^{\rm i \beta_{\omega\phi}}
}{s-m_\phi^2+im_\phi\Gamma_\phi}
\frac{s^2\Pi_\rho^{VMD}}{(1-x)s-m_\rho^2+im\rho\Gamma_\rho((1-x)s)}
\end{equation}
with $x_{1,2}=\displaystyle\frac{\mathstrut 2
p_{+,-}(p_1+p_2)}{s}$ and $x=2-x_1-x_2$. The quantities
$g^\phi_{\rho\pi}$, $g^\rho_{\pi\gamma}$ are the coupling
constants determining the $\phi\to\rho\pi$ and $\rho\to\pi\gamma$
verteces, respectively,
$F_\phi=\sqrt{\displaystyle\frac{\mathstrut 3\Gamma(\phi\to
e^+e^-)}{\alpha m_\phi}}$ and $e=\sqrt{4\pi\alpha}$. 
A fit to  the KLOE data $\phi\to\pi^0\pi^0\gamma$~\cite{kloepi0}
gives:
\begin{eqnarray}
g^\phi_{\rho\pi}=0.811{\rm GeV}^{-1} , \; \; \;
g^\rho_{\pi\gamma}&=&0.295{\rm GeV}^{-1} , \; \; \;F_\phi=42.5
,  \; \;  \Pi_\rho^{VMD}=0.58195 , \\ \nonumber \beta_\rho&=&
32.996^\circ, \; \; \beta_{\omega\phi}=163^\circ.
\end{eqnarray}

\subsection{Other contributions. 
} We included in our program the channel
$\gamma^*\to\rho^\pm\pi^\mp\to\pi^+\pi^-\gamma$~\footnote{In the
energy region $s\leq m_\phi^2$  this direct transition
$\gamma^*\to\rho^\pm\pi^\mp$ can be considered as the tail of the
double resonance contribution of the $\rho'$ meson decay:
$\gamma^*\to \rho'\to \rho\pi$ for $s=m_\phi^2$}, whose amplitude has
been evaluated in RPT model. To write this part of FSR we used the
results of Ref.~\cite{our_2005} for the function $f_i^{\rho^\pm}$:
\begin{eqnarray}
\Delta f_{1}^{\rho^\pm} &=&\frac{8 H_{V}^{2}}{9 f_{\pi }^{2}}
\biggl[ ({k\cdot Q +l^2}) \biggl( \frac{1}{C(l)}+ \frac{1}{C(-l)}
\biggr) + 2 k \cdot l \biggl( \frac{1}{C(l)} - \frac{1}{C(-l)}
\biggr) \biggr] + \frac{64 H_V^2}{9 f_\pi^2},
\label{eq:app-delta-f1} \\
\Delta f_{2}^{\rho^\pm} &=&-\frac{8 H_{V}^{2}}{9 f_{\pi }^{2}}
\biggl( \frac{1}{C(l)}+ \frac{1}{C(-l)} \biggr) ,
\label{eq:app-delta-f2} \\
\Delta f_{3}^{\rho^\pm} &=&\frac{8 H_{V}^{2}}{9 f_{\pi }^{2}}
\biggl( \frac{1}{C(l)} - \frac{1}{C(-l)} \biggr) ,
\label{eq:app-functions-in-ChPT}
\end{eqnarray}
where  $C(\pm l) =m_{\rho}^{2}- (k+ p_{\pm})^2 -im_{\rho
}\Gamma_\rho ((k+ p_{\pm})^2)$  with  $(k+ p_{\pm})^2 =
(Q^{2}+l^{2}+2k\cdot Q \pm 4k\cdot l)/4$. A value of the constant
$H_{V}$ is determined by the width of the $\rho\to\pi\gamma$ decay
$$\Gamma(\rho^\pm \to \pi^\pm \gamma) = \frac{4 \alpha m_\rho^3
H_V^2 }{27 f_\pi^2 }(1- \frac{m_\pi^2}{m_\rho^2})^3  $$  and can be related with the constant $g^\rho_{\pi\gamma}$: $H_V=\displaystyle\frac{\mathstrut 3 f_\pi g^\rho_{\pi\gamma}}{4\sqrt{2}}$. It gives $H_V=0.0144$.
 In agreement with the calculation given
in~\cite{our_2005}, we found a negligible contribution of this
channel and, for the sake of simplicity,  we discard its effects
on the numerical results presented in the following section.

\section{Numerical results}\label{numeric}
In this chapter we present the results for the differential cross section and the forward-backward asymmetry \cite{Rr2,czyz} for the reaction
$e^+e^-\to\pi^+\pi^-\gamma$, where the FSR amplitude ($M_{FSR}$)
receives contributions from both RPT ($M_{RPT}$) 
 and the $\phi\to\pi^+\pi^-\gamma$ decay  ($M_{\phi}$). The last one is a sum of the $\phi$ direct decay ($M_\phi^{scal}$) and VMD  ($M_\phi^{VMD}$) contributions.
 Thus the total contribution of the process (\ref{process}) $d\sigma_T$ 
 can be written as:
\begin{eqnarray}\label{cross_sect}
d\sigma_T & =&
d\sigma_{I}+d\sigma_{F}+d\sigma_{IF}\sim |M_{ISR}+M_{FSR} |^2 ,\\\nonumber d\sigma_{I}& \sim
& |M_{ISR}|^2, \\\nonumber d\sigma_{F}&\sim & |M_{RPT}|^2 +
|M_\phi|^2+2\mathrm{Re}\{M_{RPT}\cdot M_\phi^*\} , \\ \nonumber
d\sigma_{IF}& \sim & 2\mathrm{Re}\{M_{ISR}\cdot (M_{RPT}+
M_\phi)^*\}  \\ \nonumber
\end{eqnarray}
where $d\sigma_I$ corresponds to the ISR cross section, $d\sigma_F$ is for the FSR one.
The interference term $d\sigma_{IF}$ is equal to zero for
symmetric cuts on the polar angle of the pions~\cite{binner}.
\begin{figure}[tbp]
\label{fig1}
\par
\parbox{1.05\textwidth}{\hspace{-0.4cm}
\includegraphics[width=0.5\textwidth,height=0.5\textwidth]{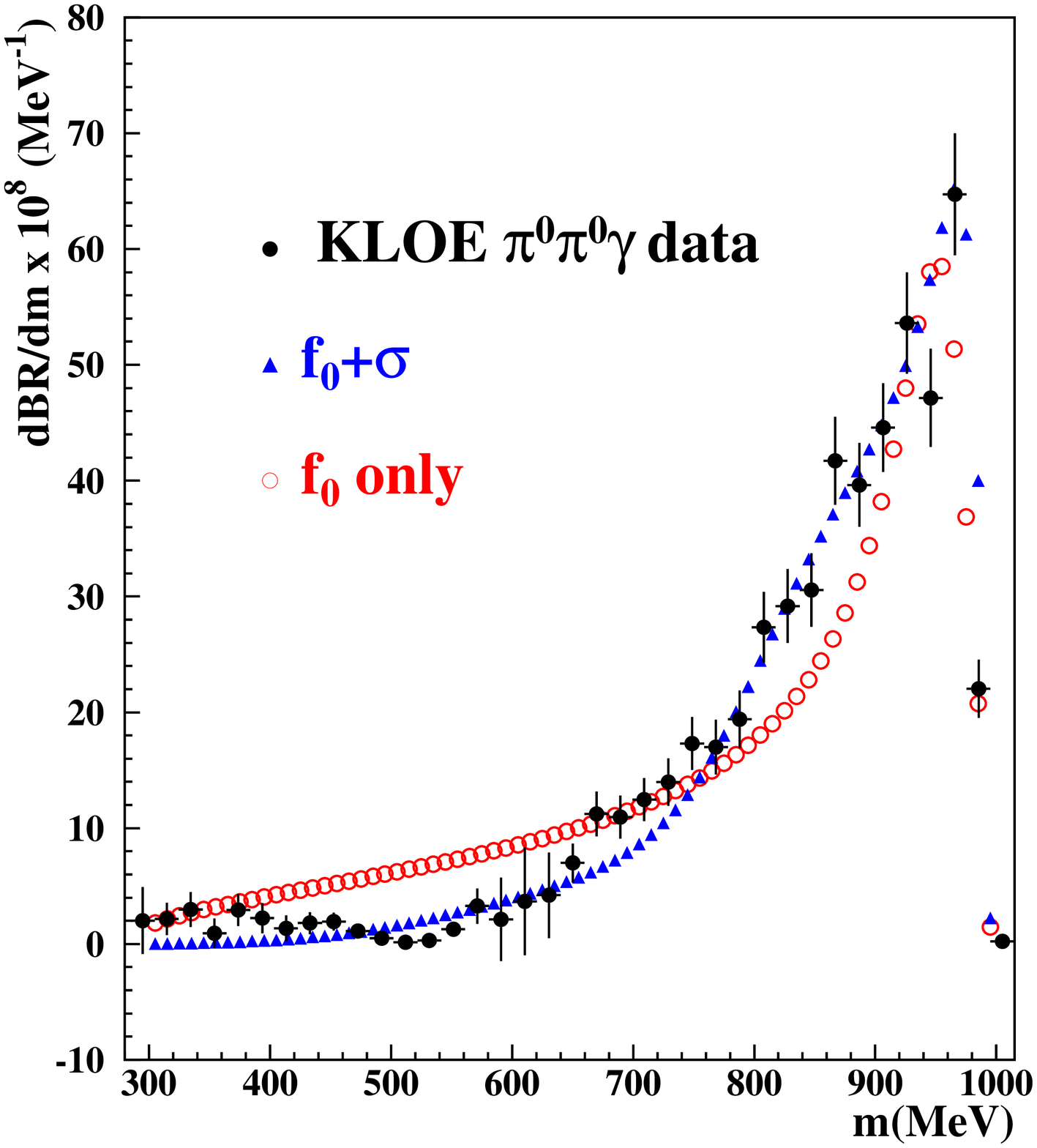}
\hspace{0.2cm}
\includegraphics[width=0.5\textwidth,height=0.5\textwidth]{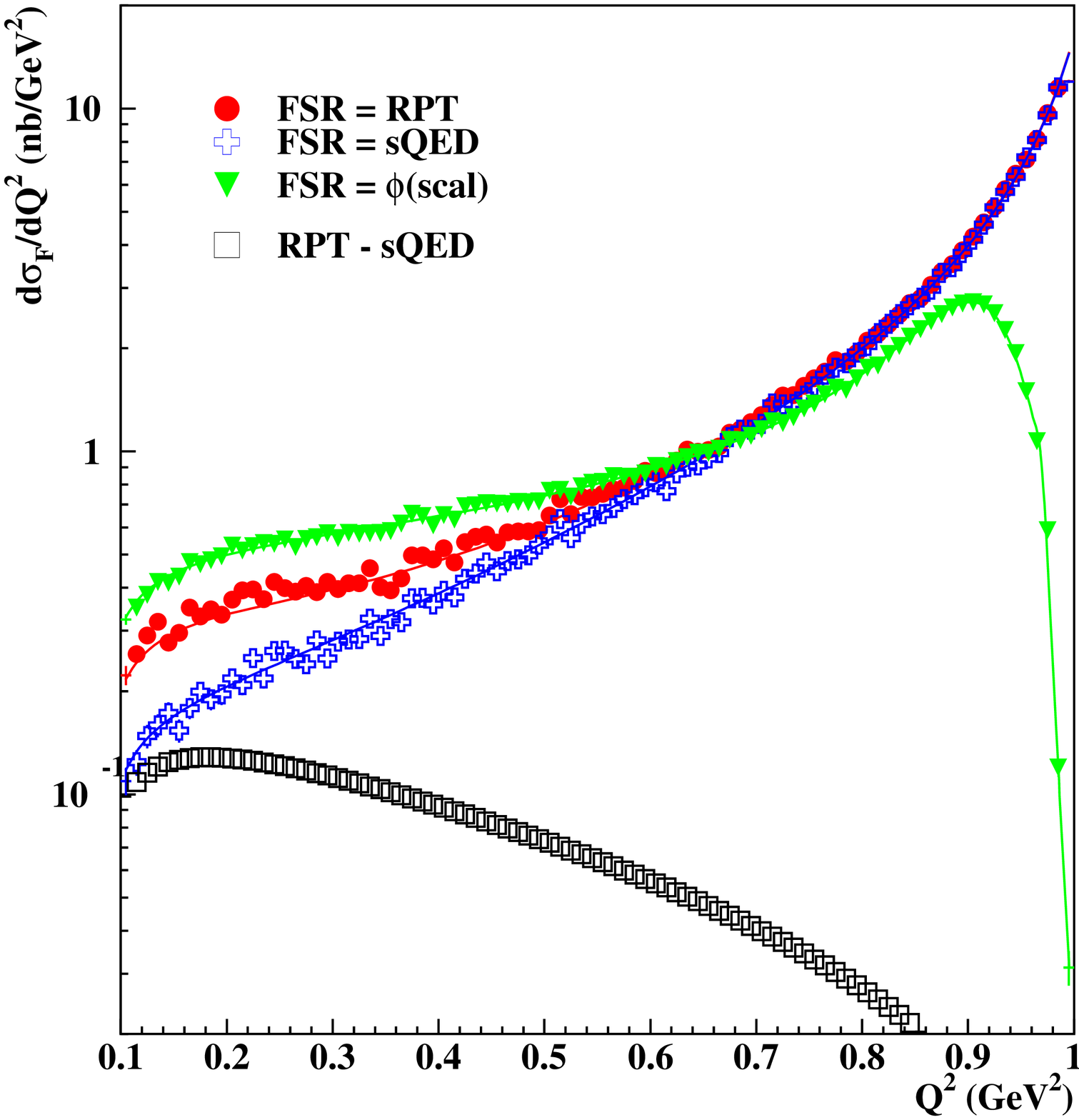}
} \caption{({\it Left}) The dependence of  the branching ratio of
the $\phi$ direct decay  on the intermediate scalar states. ({\it
Right}) Contribution to the FSR cross section $d\sigma_{F}/dQ^2$
in the region  $0^\circ\leq\theta_\gamma\leq 180^\circ$,
 $0^\circ\leq\theta_\pi\leq 180^\circ$ at $s=m_\phi^2$.
 RPT is represented by circles, sQED by crosses, $\phi$ by triangles,
while the difference between RPT and sQED is indicated by squares.
}
\end{figure}

The different contributions to the FSR differential cross section
$d\sigma_{F}$, evaluated at  $s=m_\phi^2$, are shown in Fig.1,
{\it right},
 for the full angular range $0^\circ\leq\theta_\gamma\leq 180^\circ ,
0^\circ\leq\theta_\pi\leq 180^\circ$. Good agreement between the
results of the Monte Carlo simulation (points), with the analytic
prediction (solid line) is found. It can be noted that at low
$Q^2$ the contribution from the direct $\phi$ decay (i.e. the term
proportional to $|M_\phi^{scal}|^2$ in Eq.~(\ref{cross_sect})) is
quite large
and, therefore, the additional contribution beyond sQED, can be
revealed only in the case of destructive interference between  the
two amplitudes ($\mathrm{Re}(M_{RPT}\cdot  M_\phi^*)<0$).
Published data from  the KLOE experiment~\cite{kloe} are in favour
of this assumption, which we will use in the following.

To begin with, we consider the case $s=m_\phi^2$. In Fig.2 we show
the values of the differential cross section $d\sigma_T/d\sigma_I$
and the forward-backward asymetry for the angular cuts of the KLOE
large angle analysis~\cite{kloe_large,bini}:
\begin{eqnarray}\label{angle_cuts}
50^\circ&\leq&\theta_\gamma\leq 130^\circ , \nonumber \\ 
50^\circ&\leq&\theta_\pi\leq 130^\circ
\end{eqnarray}
for the Bremsstrahlung FS process in the framework of the sQED$*$VMD model  and with the  $\phi$  decay contributions (VMD and the $\phi$ direct decay), for a  hard photon radiation with
energies $E_\gamma>20$ MeV.
\begin{figure}
\label{fig2}
\par
\parbox{1.05\textwidth}{\hspace{-0.4cm}
\includegraphics[width=0.5\textwidth,height=0.5\textwidth]{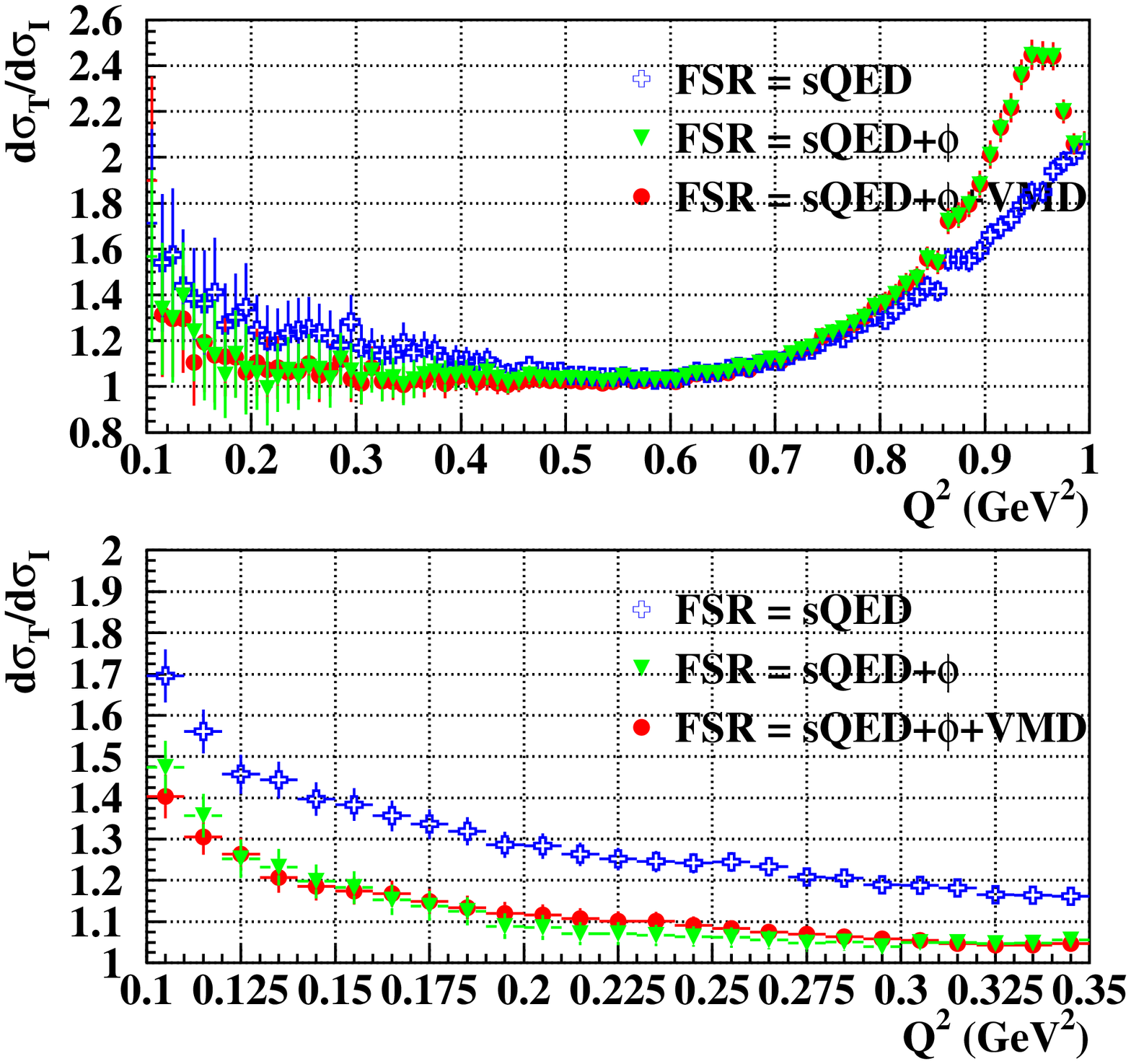}\hspace{0.4cm}
\includegraphics[width=0.5\textwidth,height=0.5\textwidth]{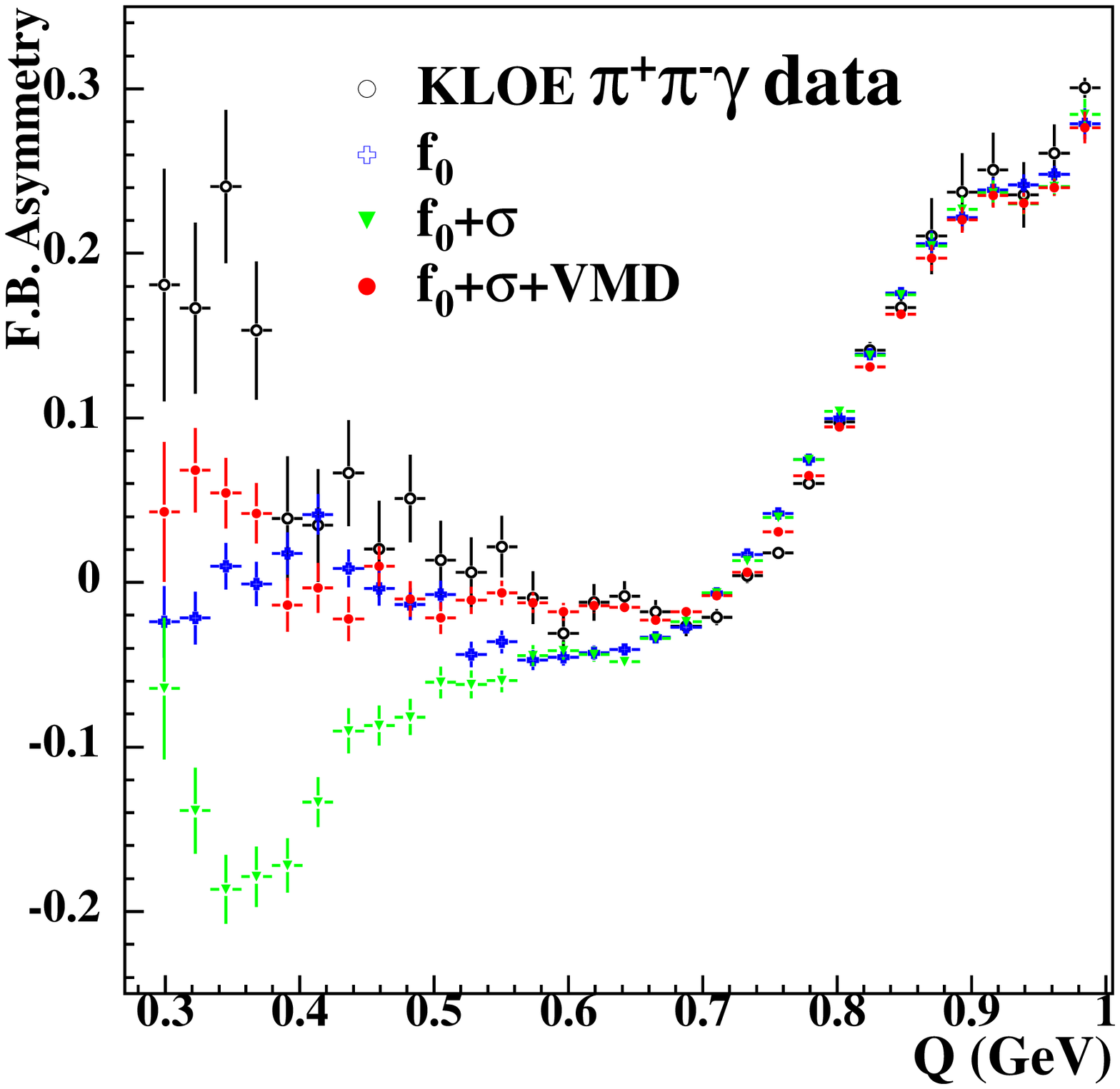}}
\vspace{-0.5cm}
\caption{
The ratio of the total cross section  with respect to ISR one
$d\sigma_T/d\sigma_I$ ({\it left}) and the forward-backward
asymmetry  ({\it right}) as function of the invariant mass of the
two pions, when the $\phi$ contribution is taken into account and
the Bremsstrahlung process is in the framework of the sQED$*$VMD
model. The angular region is $50^\circ\leq\theta_\gamma\leq
130^\circ$, $50^\circ\leq\theta_\pi\leq 130^\circ$ and
$s=m_\phi^2$. }
\end{figure}

\begin{figure}[tbp]
\label{fig3}
\par
\parbox{1.05\textwidth}{\hspace{-0.4cm}
\includegraphics[width=0.5\textwidth,height=0.5\textwidth]{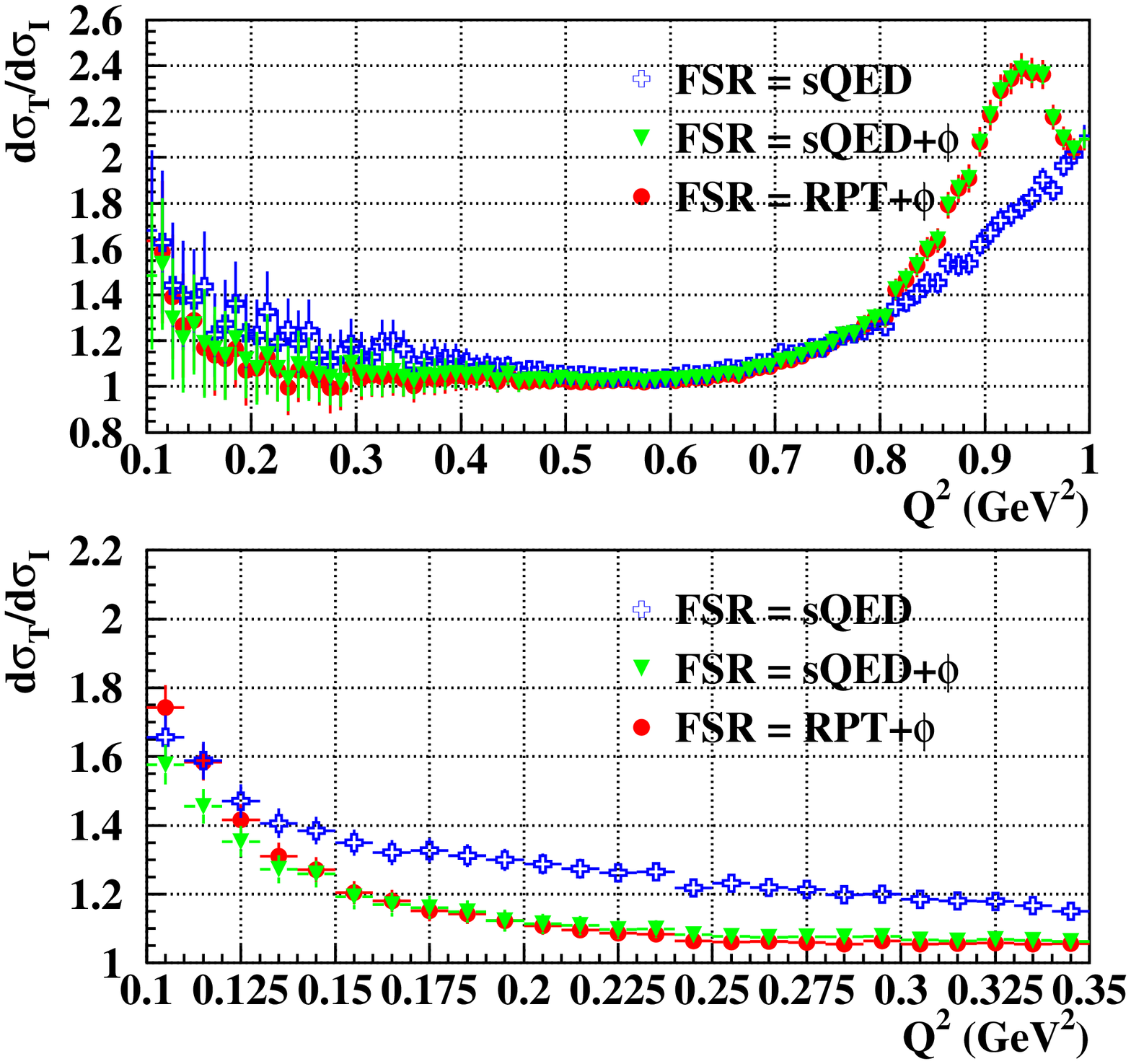}
\hspace{0.2cm}
\includegraphics[width=0.5\textwidth,height=0.5\textwidth]{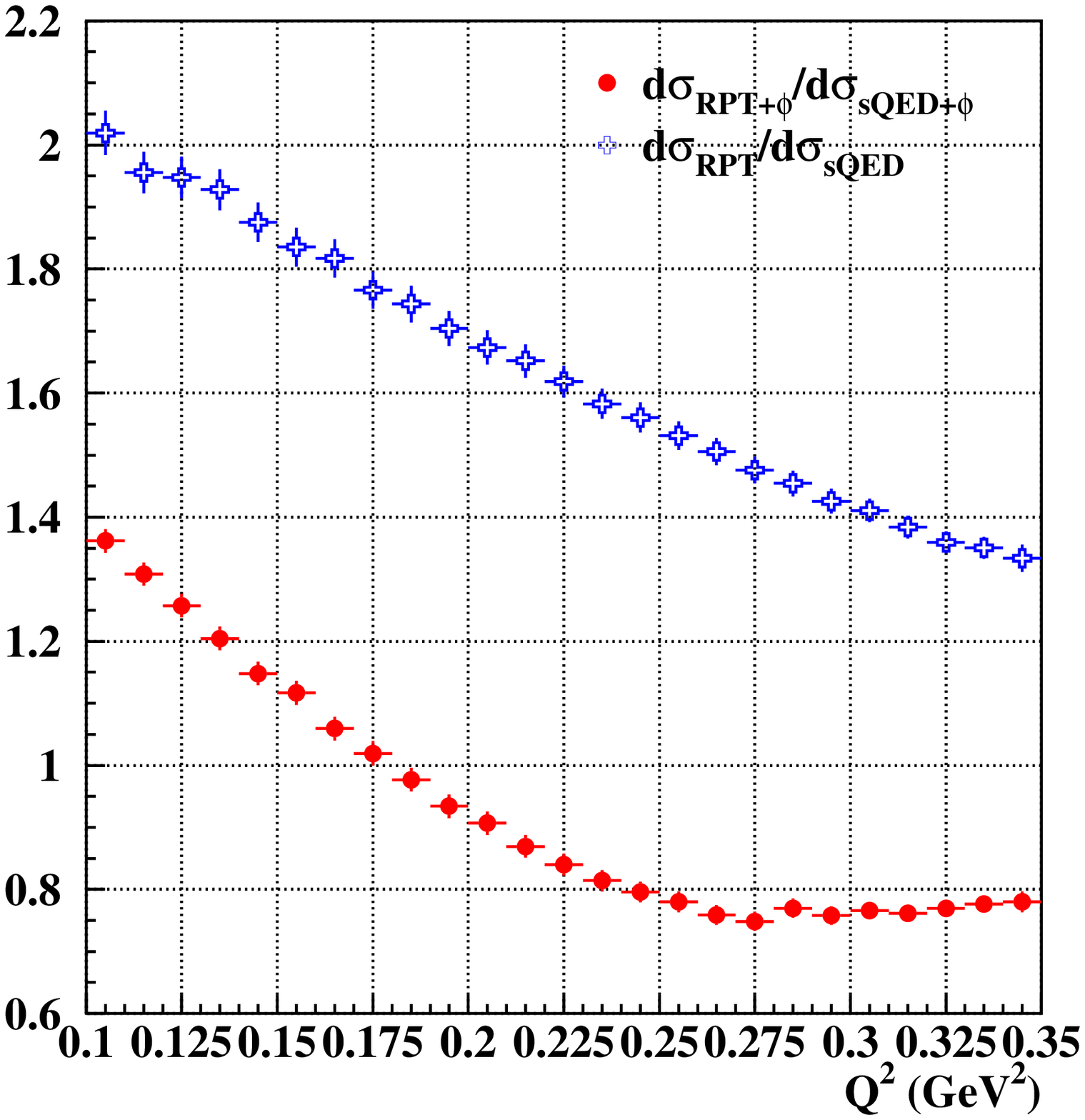}}
\caption{The ratio $d\sigma_T/d\sigma_I$ ({\it left}) as function
of the invariant mass of the two pions for  different models
describing the Bremsstrahlung FS process (either RPT or sQED$*$VMD)
and the  ratio of FSR cross section in the framework of RPT, with
respect to sQED, when the $\phi$ direct decay contribution is (or
not) taken into account ({\it right}), in the region
$50^\circ\leq\theta_\gamma\leq 130^\circ$,
$50^\circ\leq\theta_\pi\leq 130^\circ$ at  $s=m_\phi^2$.  }
\end{figure}

Figure~3 shows the effects of RPT and $\phi$ direct decay terms to
the total differential cross section  and their contribution to FSR one for angular cuts (\ref{angle_cuts}). 
Several distinctive features can be noted:{\it (1)} the peak at
about $1$ GeV$^2$ corresponds to the $f_0$ intermediate state for
the direct $\phi\to\pi\pi\gamma$ amplitude; {\it (2)} the presence
of RPT terms in the FSR is relevant at low $Q^2$, where they give
 an additional contribution up to $40\%$ to the ratio
$d\sigma_{RPT+\phi}/d\sigma_{sQED+\phi}$; (as shown in Fig.~3,{\it left});
{\it (3)}
the destructive interference between the $\phi$ decay an the
Bremsstrahlung FS process decreases the visible cross section in
the whole $Q^2$ region and its dependence on FS Bremsstrahlung
model at low $Q^2$ (see Fig.~2, {\it left}, down). Also we would
like to draw attention to the VMD contribution. As we can see from
Fig.~2, the VMD contribution almost does not change the value of
the differential cross section (Fig.~2, {\it left}), but it
changes essentially the value of the forward-backward asymmetry
(Fig.~2, {\it right}) and it follows the experimental data for it.
Last but not least, all contributions beyond  sQED are large
enough near the threshold to make the analysis difficult.

\begin{figure}[tbp]
\label{fig4}
\par
\parbox{1.05\textwidth}{\hspace{-0.4cm}
\includegraphics[width=0.5\textwidth,height=0.5\textwidth]{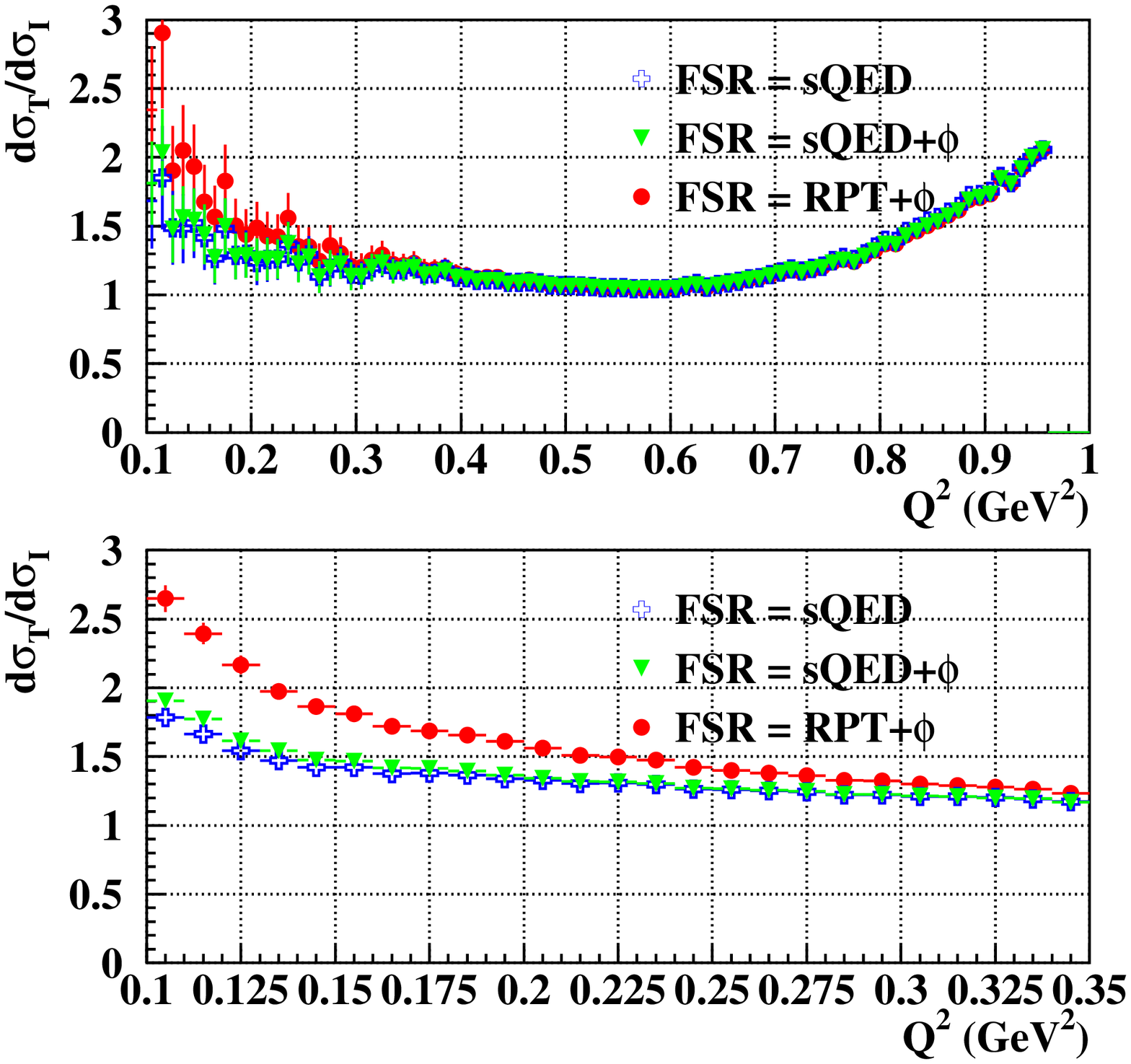}
\hspace{0.2cm}
\includegraphics[width=0.5\textwidth,height=0.5\textwidth]{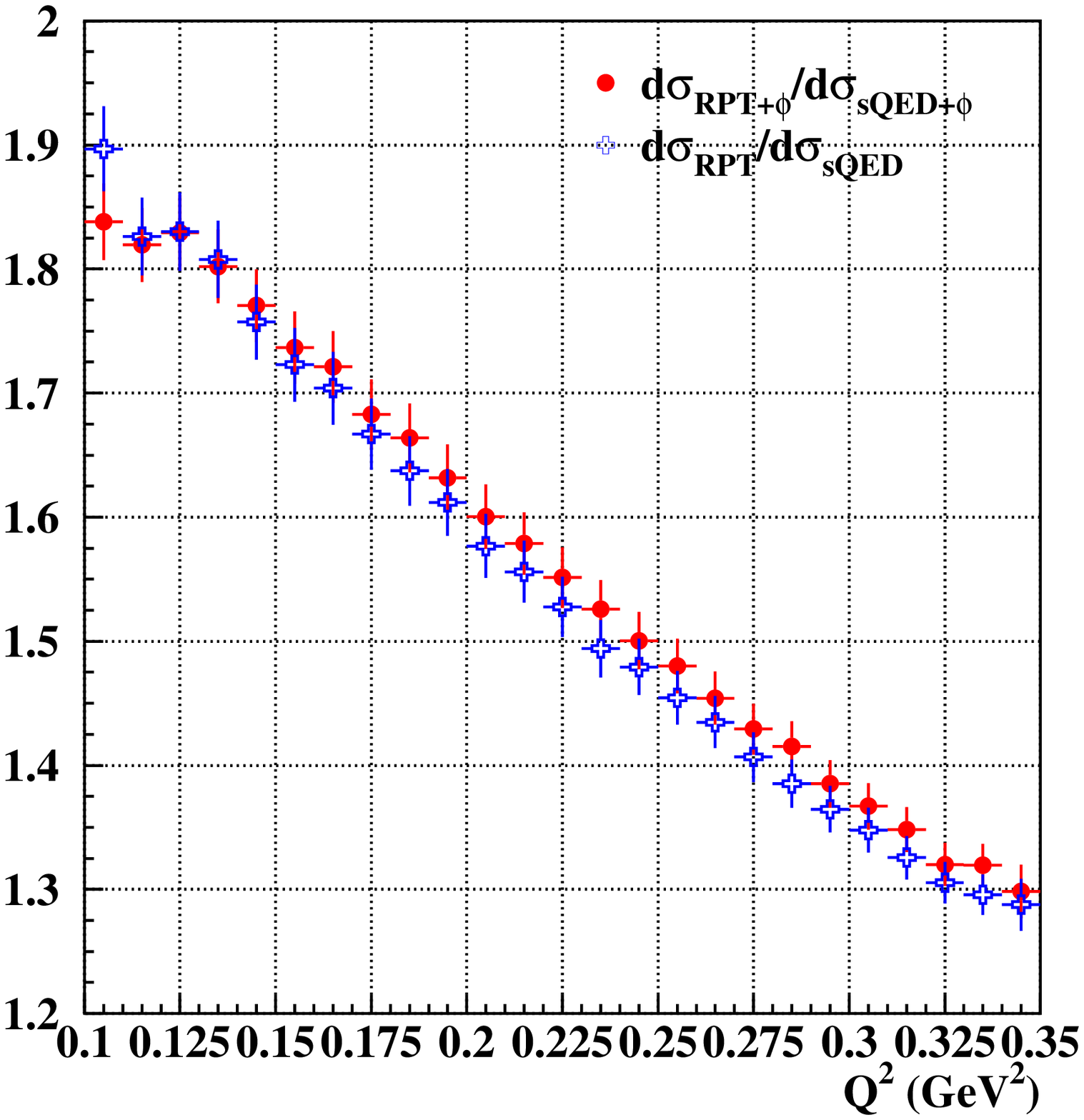}}
\caption{The same ratio as in Fig.~3 at  $s=1$GeV$^2$. }
\end{figure}

In order to reduce the background from  $\phi$ decay in the
measurement of the pion form factor at threshold, KLOE has taken
more than $200$ pb$^{-1}$ of data at 1 GeV~\cite{stef}. In this
case the $\phi$ meson intermediate contributions
are suppressed (see Fig. 4, (right), the values of $d\sigma_T$ with and
without the $\phi$  decay almost coincide) and the main contribution to FSR comes from
Bremsstrahlung process (see Section \ref{Br_proc}), allowing to study a model for it.


\section{Model-independent test of FSR models}\label{mod_ind}
Contributions to Bremsstrahlung FS process beyond sQED, as in the
case of RPT, can lead to sizeable effects on the cross section and
asymmetry at threshold, as shown in Figs.~2-4. Precise
measurement of the pion form factor in this region needs to
control them at the required level of accuracy. This looks like a
rather difficult task, if one thinks that  effects
 beyond sQED, as well as the contribution from $\phi\to\pi^+\pi^-\gamma$,
are model-dependent.

One can think to construct a general amplitude for the
$e^+e^-\to\pi^+\pi^-\gamma$, according to some underlying theory,
and try to determine the free parameters by a constrained fit on
specific variables (like mass spectrum, charge and
forward-backward asymmetry, angular distribution, etc...).
Particularly for the charge asymmetry, it has
 been proved to be a powerful tool to discriminate between
different models of  $\phi\to\pi^+\pi^-\gamma$~\cite{czyz}.
 However, when the number of the parameters is large,
correlations between the parameters of the model can arise and
spoil the effective power of these fits. The situation becomes
even worse if the pion form factor also has to be extracted from
the same data. As an example, in the case of RPT model,  if we
consider only the $\rho$ and $\omega$ contribution to the pion
form factor and the $\rho$ and $a_1$ contribution to FSR the
number of free parameters is already six. The presence of the
$\phi$ direct
 and VMD decays results in additional free parameters.

The possibility to determine some of the parameters by external
data can strongly help, as in the case of the
$\phi\to\pi^+\pi^-\gamma$ amplitude, which can be determined by
the $\pi^0\pi^0\gamma$ channel copiously produced at DA$\Phi$NE.
An additional source of information, which will be used to
determine the contributions to FSR beyond sQED in a
model-dependent way, is the dependence of the FSR amplitude
on the $e^+e^-$ invariant mass squared $s$.

Let us write the differential cross section for the emission of
one photon in
 the process $e^+e^-\to\pi^+\pi^-\gamma$ as a function of the
invariant mass of the two pions:
\begin{equation}\label{spectr_low}
\Big(\frac{d\sigma_T}{dQ^2}\Big)_s= |F_\pi(Q^2)|^2
H_s(Q^2)+\Big(\frac{d\sigma_F}{dQ^2}\Big)_s,
\end{equation}
where $\Big(\mathstrut\frac{\displaystyle d\sigma_F}{dQ^2}\Big)_s$
is the differential cross section for the emission of a photon in
the final state, while the ISR function  $H_s(Q^2)$ was defined in the Introduction. 
We indicate by a subscript $s$  the dependence of each quantity on
the $e^+e^-$ invariant mass ($s$). Since we will consider only
symmetric angular cuts for pions, the  interference term between
initial and final state radiation has been neglected.

At relatively high $Q^2$ the FSR differential cross section,
$\Big(\mathstrut\frac{\displaystyle d\sigma_F}{dQ^2}\Big)_s$, is
dominated  by the contribution coming from sQED ($M_{sQED}$) and
$\phi$ direct decay ($M_{\phi}$): \be
\Big(\frac{d\sigma_{sQED+\phi}}{dQ^2}\Big)_s \sim  |M_{sQED}+
M_{\phi} |^2 . \ee Contributions beyond sQED ($\Delta M$) are
expected to be important at low $Q^2$. They introduce  an
additional term ($\Delta M$) in the above expression:
\begin{eqnarray}
\Big(\frac{d\sigma_F}{dQ^2}\Big)_s  &\sim &| M_{sQED}+ \Delta M + M_{\phi} |^2 = \\
& = & |M_{sQED}+ M_{\phi} |^2 + |\Delta M|^2 +
2\mathrm{Re}\Big\{\Delta M \cdot (M_{sQED}+M_\phi)^* \Big\} .
\end{eqnarray}


We will now consider the following quantity:
\begin{equation}\label{y}
Y_s(Q^2)=\frac{ \Big (\frac{d\sigma_T}{dQ^2}\Big)_s-
\Big(\frac{d\sigma_{sQED+\phi}}{dQ^2}\Big)_s}{H_s(Q^2)} =
|F_\pi(Q^2)|^2+\Delta F_s(Q^2),
\end{equation}
where $\Delta F_s \sim \Big(|\Delta M|^2 +
2\mathrm{Re}\Big\{\Delta M \cdot (M_{sQED}+M_\phi)^*
\Big\}\Big)_s/ H_s$.

If no contribution beyond sQED is present ($\Delta M = 0$),
$Y_s(Q^2)$ coincides with the square of the pion form factor,
\textbf{independently of the energy $\sqrt{s}$ at which it is
evaluated}, while any dependence on $s$ is only due to an
additional contribution to FSR beyond sQED.
In  particular, the difference of $Y_s(Q^2)$ computed at two beam
energies ($s_1$ and $s_2$), can only come from FSR beyond sQED:
\begin{equation}
\Delta Y(Q^2) = Y_{s_1}(Q^2)-Y_{s_2}(Q^2) =  \Delta F_{s_1}(Q^2)-
\Delta F_{s_2}(Q^2)
\end{equation}
Therefore, before extracting the pion form factor at threshold, we
suggest to look at the difference
 $\Delta Y(Q^2)$, which  can be used to estimate the
contribution beyond sQED to the FSR amplitude in a model-independent
way.


As a realistic application of this procedure, we consider the case
of DA$\Phi$NE, where KLOE has already collected more than 200
pb$^{-1}$ at 1 GeV$^2$ and 2.5 fb$^{-1}$ at $m_\phi^2$, that, in
the range $Q^2<0.35$ GeV$^2$ , correspond to $O(10^3)$ and
$O(10^4)$ events, respectively, in the region
$50^\circ\leq\theta_\gamma\leq 130^\circ$,
$50^\circ\leq\theta_\pi\leq 130^\circ$. We will consider RPT as a
model for the effects beyond sQED.
\begin{figure}
\label{fig5}
\par
\parbox{1.05\textwidth}{\hspace{-0.3cm}
\includegraphics[width=0.5\textwidth,height=0.5\textwidth]{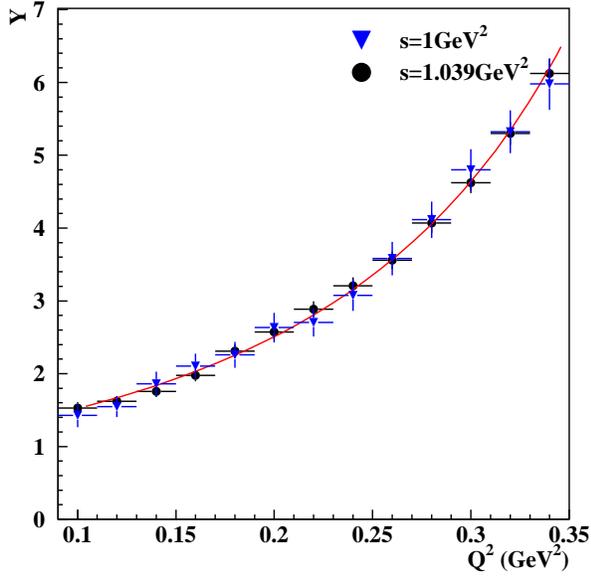}
\hspace{0.3cm}
\includegraphics[width=0.5\textwidth,height=0.5\textwidth]{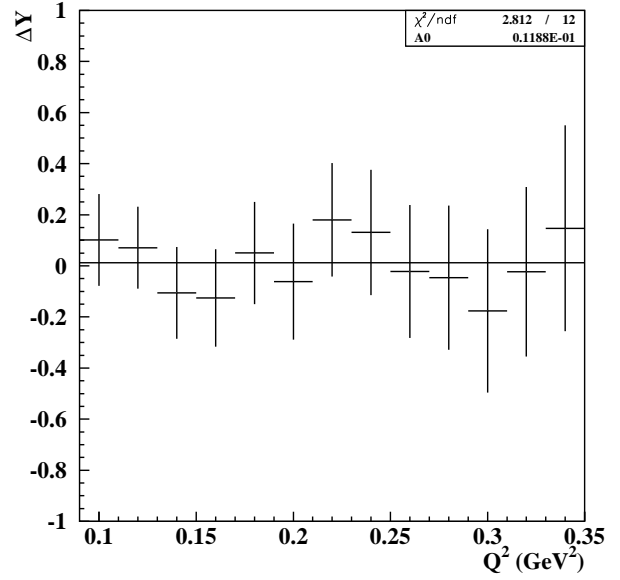}}
\caption{ {\it Left}: $Y_s(Q^2)$ at $s=1$ GeV$^2$ (triangles),
 and at $s=m_\phi^2$  (circles), when FSR includes only the sQED and
$\phi$ contribution.  The  pion form factor $|F_{\pi}(Q^2)|^2$ is
shown by a solid line. {\it Right}: The difference $\Delta
Y(Q^2)$. }
\end{figure}
\begin{figure}
\label{fig6}
\par
\parbox{1.05\textwidth}{\hspace{-0.3cm}
\includegraphics[width=0.5\textwidth,height=0.5\textwidth]{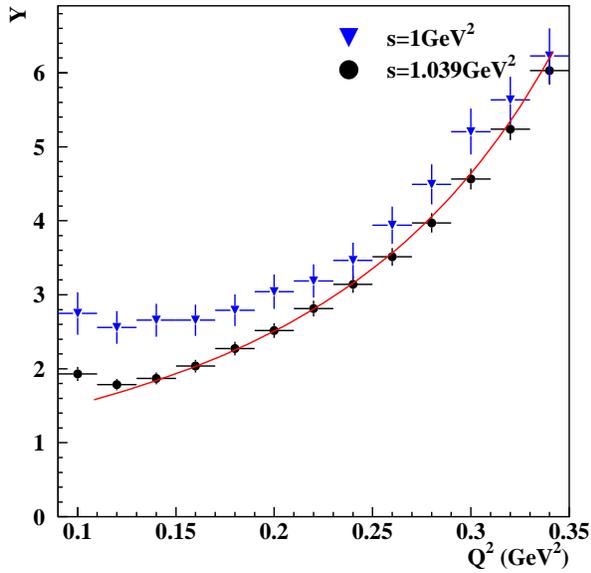}
\hspace{0.3cm}
\includegraphics[width=0.5\textwidth,height=0.5\textwidth]{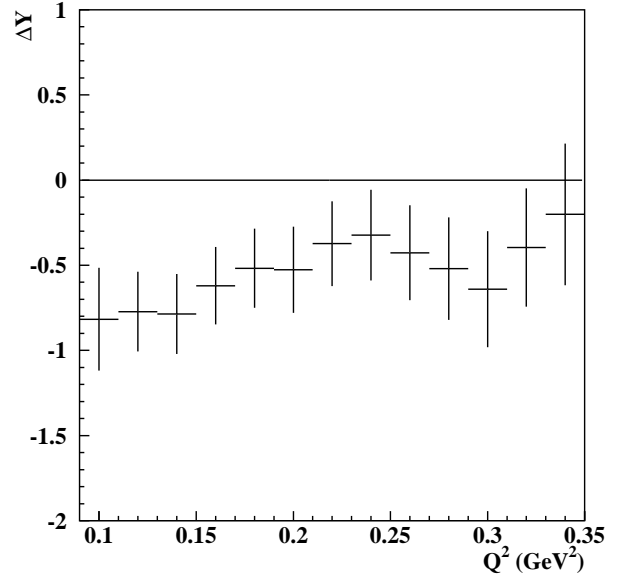}}
\caption{
{\it Left}: $Y_s(Q^2)$ at $s=1$ GeV$^2$ (triangles),
 and at $s=m_\phi^2$ (circles), when FSR includes RPT and $\phi$
contribution. The pion form factor $|F_{\pi}(Q^2)|^2$ is shown by
a solid line. {\it Right}: The difference $\Delta Y(Q^2)$. }
\end{figure}

Fig. 5, {\it left}, shows the quantity $Y_s(Q^2)$ at $s_1=1$
GeV$^2$ and at $s_2=m_\phi^2$
 when  no additional RPT term is included in FSR.
As expected, each of these quantities coincides with the square of
the pion form factor $|F_{\pi}(Q^2)|^2$, shown by a solid line.
The difference $\Delta Y(Q^2)$ is shown in Fig.5, {\it right},
which is consistent with zero as expected. In the region $s<0.35$
GeV$^2$ one can expand the pion form factor  as it was done
in~\cite{eid_jeg}: \begin{equation}\label{decomp_fpi}
F_\pi(q^2)\simeq 1+ p_1\cdot q^2+p_2\cdot q^4 .
\end{equation}
Using the same experimental data for the pion form factor \cite{cmd2} as before we have: $p_1=1.15\pm 0.06$ GeV$^{-2}$,
$p_2=9.06\pm 0.25$ GeV$^{-4}$, $\chi^2/\nu\simeq
0.13$. A combined fit of $Y_s(Q^2)$ to the pion form factor  
gives the following values:
 $p_1=1.4\pm 0.186$ GeV$^{-2}$, $p_2=8.8\pm 0.73$ GeV$^{-4}$,
 $\chi^2/\nu=0.25$, that is
  in a reasonable agreement with the results (\ref{decomp_fpi}).

The situation is different as soon as the Bremsstrahlung FS process is
modeled by RPT. In this case, as shown in Fig.6, {\it right}, the
difference $\Delta Y(Q^2) \neq 0$ and the quantities $Y_s(Q^2)$
cannot be anymore identified with $|F_{\pi}(Q^2)|^2$, (see Fig.6,
{\it left})~\footnote{Destructive interference between RPT and
 $\phi\to\pi^+\pi^-\gamma$ amplitudes tends to cancel out the effects beyond sQED at $s=m_\phi^2$ (see Fig.~3, {\it left}). Therefore the quantity $Y_s(Q^2)$ almost coincides with the pion form factor.}. 


Before concluding, we would like to point out the main points
of our present method:

\begin{itemize}

\item The quantity  $\Big(\frac{d\sigma_{sQED+\phi}}{dQ^2}\Big)_s$
is an input parameter of our procedure, and can be computed
numerically by Monte Carlo;

\item The amplitude for $\phi\to\pi^+\pi^-\gamma$
is taken from the  $\pi^0\pi^0\gamma$ channel.

\item missing ISR multi-photon radiative correction can be added  in $H_s$ and it will not spoil the effective power of the method;

\item A clear advantage of the procedure based on
a Monte Carlo event generator is that it allows to keep control
over efficiency and resolution of the detector and fine tuning of
the parameters.

\end{itemize}

Even if  the main limitation of the method could come from the
uncertainty on the parameters of $\phi\to\pi^+\pi^-\gamma$
amplitude, especially at low $Q^2$, we believe that the KLOE data
on $\phi\to\pi^0\pi^0\gamma$  will allow a precise
description of this amplitude. In any case, in agreement
with~\cite{czyz} we strongly recommend to check the amplitude by
using
 charge asymmetry and to compare with
spectrum of the $\pi^+\pi^-\gamma$, at least at high $Q^2$, where
the pointlike approximation is safe (as done in~\cite{bini}).

\section{Conclusion}\label{conclusion}
A  test of FSR at threshold in the process
$e^+e^-\to\pi^+\pi^-\gamma$
 is a rather important issue, not only for the role of FSR as background to
the measurement of the pion form factor, but also to get
information about pion-photon interactions when the intermediate
hadrons are far off shell. At $s=m_{\phi}^2$ an additional
complication arises: the presence of the  decay
$\phi\to\pi^+\pi^-\gamma$, that goes either through the
intermediate scalar (the direct $\phi$ decay) or the vector state
(VMD contribution) whose amplitude and relative phase can be
described according to some model. By means of a Monte Carlo event
generator FEVA, which also includes the contribution of the rare
$\phi\to\pi^+\pi^-\gamma$ decay, we estimate the effects beyond
sQED in the framework of Resonance Perturbation Theory (RPT)
for angular cuts used in the KLOE analysis of the pion form factor
at threshold. We show that the low $Q^2$ region is sensitive both
to the inclusion of additional terms in the FSR amplitude given by
the RPT model and to the $\phi$ decay contribution (especially its
VMD part).

We also propose a method
which allows to estimate the effects beyond sQED in a
model-independent way. We found that the deviation from sQED
predicted by RPT can be observed within the current KLOE statistics.

We would like to emphasize once again that this work was motivated
by  ongoing experiment on precise measurements of the muon anomalous
magnetic moment \cite{amu_fut} that allows to perform tests of the Standard
Model with a fabolous precision.

\vspace{1cm}

\textbf{Acknowledgements}
It is pleasure to thank all our colleagues of the 'Working Group on Radiative Corrections and MC Generators for Low Energies' for many useful discussions.
 We are especially grateful to S. Eidelman  for useful discussion and careful reading of the manuscript. This work has been supported by the grant INTAS/05-1000008-8328. Also G.P. acknowledges  support from EU-CT2002-311 Euridice contract. 


\end{document}